\documentclass[prl,reprint,superscriptaddress,amsmath,amssymb,floatfix,showpacs]{revtex4-1}

\usepackage{hyperref}
\usepackage{graphicx,graphics}

\begin{document}

\newcommand{\IEF}{Institut d'Electronique Fondamentale, UMR CNRS 8622, Univ. Paris-Sud, 91405 Orsay, France}
\newcommand{\UMPhy}{Unit\'e Mixte de Physique CNRS/Thales, 1 avenue A. Fresnel, Campus de l'Ecole Polytechnique, 91767 Palaiseau, France,  and Univ. Paris-Sud, 91405 Orsay, France}
\newcommand{\UFRGS}{Instituto de Física, Universidade Federal do Rio Grande do Sul, Porto Alegre 91501-970, Brazil}
\newcommand{\AIST}{National Institute of Advanced Industrial Science and Technology (AIST) 1-1-1 Umezono, Tsukuba, Ibaraki 305-8568, Japan}

\title{Noise-enhanced synchronization of stochastic magnetic oscillators}

\author{N. Locatelli}
\author{A. Mizrahi}
\affiliation{\UMPhy}
\affiliation{\IEF}

\author{A. Accioly}
\affiliation{\UMPhy}
\affiliation{\IEF}
\affiliation{\UFRGS}

\author{R. Matsumoto}
\author{A. Fukushima}
\author{H. Kubota}
\author{S. Yuasa}
\affiliation{\AIST}

\author{V. Cros}
\affiliation{\UMPhy}

\author{L. G. Pereira}
\affiliation{\UFRGS}

\author{D. Querlioz}
\author{J.-V. Kim}
\affiliation{\IEF}

\author{J. Grollier}
\email{julie.grollier@thalesgroup.com}
\affiliation{\UMPhy}

\begin{abstract}
We present an experimental study of phase-locking in a stochastic magnetic oscillator. The system comprises a magnetic tunnel junction with a superparamagnetic free layer, whose magnetization dynamics is driven with spin torques through an external periodic driving current. We show that synchronization of this stochastic oscillator to the input current is possible for current densities below $3 \times 10^6$ A/cm$^2$, and occurs for input frequencies lower than the natural mean frequency of the stochastic oscillator. We show that such injection-locking is robust and leads to a drastic reduction in the phase diffusion of the stochastic oscillator, despite the presence of a frequency mismatch between the oscillator and the excitation.  
\end{abstract}

\maketitle


Spin torque driven magnetic tunnel junctions exhibit a variety of dynamic behaviors, which combined with their tiny size, CMOS compatibility and endurance, make them promising candidates for a range of applications~\cite{NM_13_Locatelli_2014}. In particular, they are ideal model systems for the study of nonlinear dynamical phenomena. Under certain conditions, spin torques can lead to self-sustained precession of the magnetization and the junctions behave as nonlinear auto-oscillators, which can phase-lock to an external signal~\cite{PRL_95_Rippard_2005, PRL_101_Georges_2008, PRL_105_Urazhdin_2010, APL_98_Dussaux_2011}, or even self-synchronize \cite{N_437_Kaka_2005, N_437_Mancoff_2005, PRB_73_Grollier_2006}. Phenomena related to the latter are currently the focus of much research, as it represents a promising means of improving the quality factor of such spin torque nano-oscillators. However, the experimental demonstration of self-synchronization has until now been limited to a small number of spin torque nano-oscillators (N $\leq$ 4) \cite{NN_4_Ruotolo_2009, NC_4_Sani_2013}.

Due to the small magnetic volume (or ``active region'') of such oscillators, the magnetization dynamics of these nano-objects is very sensitive to thermal fluctuations and other noise sources, resulting in a large phase noise that is detrimental to efficient phase-locking and synchronization~\cite{APL_97_Quinsat_2010, PRB_89_Grimaldi_2014}. To make progress towards synchronization, one line of enquiry has involved studying modes that are less sensitive to noise, such as vortex gyration~\cite{PRL_100_Mistral_2008, NP_3_Pribiag_2007, NC_1_Dussaux_2010} or excitations in coupled bilayers \cite{APL_98_Locatelli_2011, APL_99_Gusakova_2011}. Here, we instead pursue a different paradigm in which noise can be advantageous for improving coherence and facilitating synchronized states. This builds upon recent work in which spin torque driven magnetic tunnel junctions have been shown to exhibit stochastic resonance \cite{PRL_105_Cheng_2010, APL_103_Cheng_2013, PRB_83_Finocchio_2011}, i.e., a noise-enhanced sensitivity to very weak external stimuli \cite{RMP_70_Gammaitoni_1998}. Our work is focused on the \emph{stochastic oscillator}, which involves a bistable state in which thermal fluctuations drive random transitions between the two states. Because such transitions involve a mean transition rate, an average frequency for the stochastic oscillator can be defined, and the presence of thermal fluctuations means that switching between the two states persists indefinitely without any external forcing. 

\begin{figure}[b!]
	\centering
	\includegraphics[width=\linewidth]{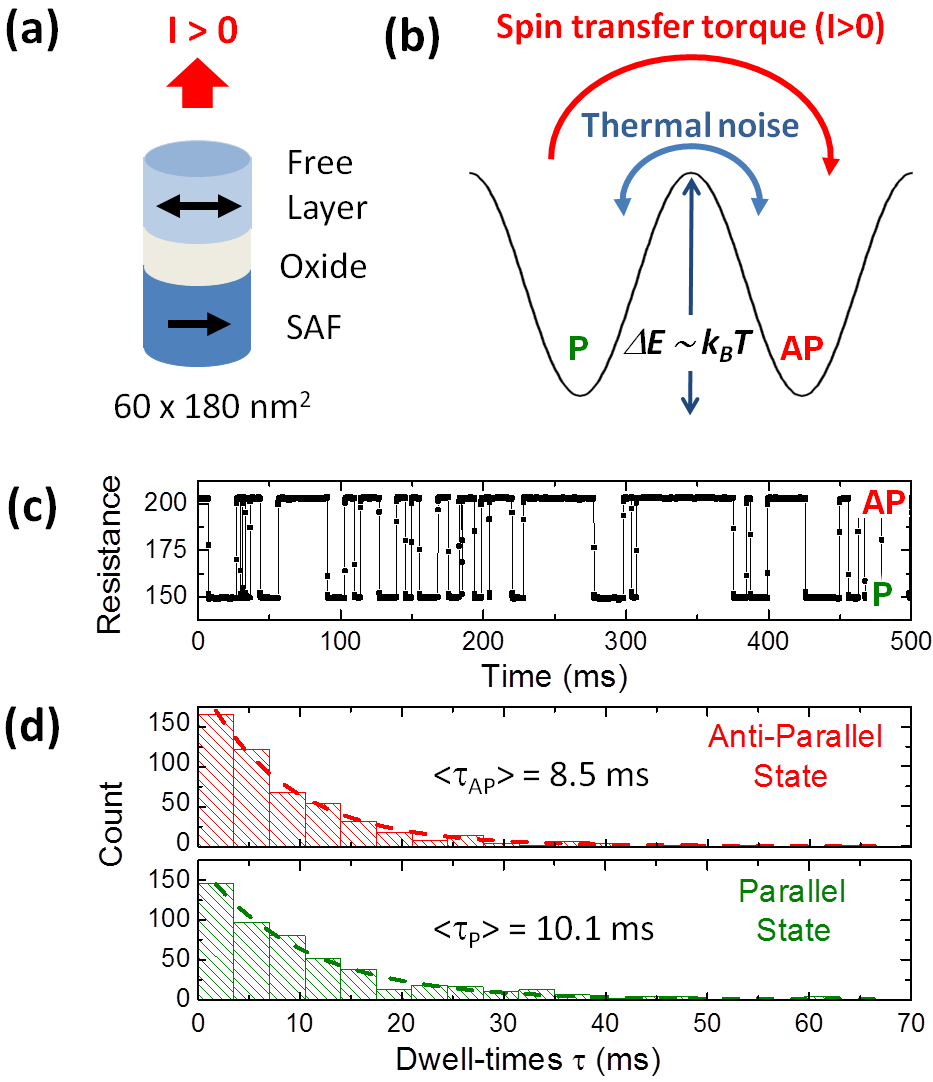}
	\caption{(a) Schematic of the magnetic tunnel junction stack. (b) Schematic of double well energy landscape for the free layer magnetization, that can hop between the P and AP states with the assistance of thermal noise and spin torque. (c) Sample of the telegraphic temporal resistance/magnetization evolution generated by the MTJ, measured at room temperature under a $I_{\rm dc}=100\mu$A current and a $10$Oe bias field. (d) Associated dwell-times distributions for P and AP states, fitted by an exponential envelope corresponding to Poissonian distribution.}
	\label{fig:SamplePres}
\end{figure}

Theoretical studies have shown that synchronization of such bistable stochastic oscillators is possible with an external harmonic excitation~\cite{PRE_58_Neiman_1998, CWN_13_Freund_2003}.  In this Letter, we demonstrate experimentally that noise can be leveraged to phase-lock stochastic magnetic oscillators at a low energy cost. Our system is a superparamagnetic tunnel junction: at zero or small bias, the free layer magnetization constantly fluctuates between the parallel (P) and antiparallel (AP) states under thermal fluctuations. This system is particularly interesting for two main reasons. First, the energy barrier between the P and AP states is low, which allows the magnetic configuration to be manipulated with spin torques at very small current densities. Second, the junction naturally behaves as a stochastic oscillator as described above. We show using time-resolved experiments that, despite its stochastic nature, phase-synchronization of the superparamagnetic tunnel junction can be achieved by using low densities for the ac excitation current ($< 3 \times 10^6$ A/cm$^2$) compared to those usually required for spin torque nano-oscillators~\cite{PRL_95_Rippard_2005, PRL_101_Georges_2008, PRL_105_Urazhdin_2010, APL_98_Dussaux_2011}. By investigating the phase diffusion processes at stake we demonstrate the existence of a critical frequency under which the noisy system is able to lock to the excitation.

The stochastic oscillator is a magnetic tunnel junction (MTJ) with an elliptical cross-section of $60\times180$ nm$^{2}$, composed of a reference synthetic antiferromagnetic (SAF) trilayer of CoFe (2.5 nm)/Ru (0.85 nm)/CoFeB (3 nm), an MgO tunnel barrier (1.05 nm), and a CoFeTiB (2 nm) free layer (Fig.~\ref{fig:SamplePres}a). The free layer magnetization is bistable, where the energy barrier separating the two states is designed to be sufficiently low such thermally-induced switching of the free layer magnetization from one state to the other occurs at room temperature (Fig.~\ref{fig:SamplePres}b). The MTJ is then said to be superparamagnetic (SP-MTJ) and oscillates stochastically between the P and AP states, generating a telegraph resistance signal due to the large tunneling magnetoresistance (TMR) of the junction (Fig.~\ref{fig:SamplePres}c). 
Under constant bias, the dwell times in each state follow a Poisson distribution (Fig.~\ref{fig:SamplePres}d)~\cite{PRB_84_Rippard_2011}. For instance, we measure at low currents ($I_{\rm dc}=100$ $\mu$A) mean dwell-times of $\langle\tau_{\rm AP}\rangle=10.1$ ms in the high resistance AP-state and $\langle\tau_{\rm P}\rangle=8.5$ ms  in the low resistance P-state. In this ``free-running'' regime, the junction behaves as a stochastic oscillator and can be characterized by its mean frequency, defined by $\langle F \rangle=1/\left(\langle\tau_{\rm P}\rangle+\langle\tau_{\rm AP}\rangle\right)$, which corresponds to $\langle F \rangle=53.8$ Hz here. As illustrated in Fig.~\ref{fig:SamplePres}(b), depending on the sign of the injected current, the spin torque will favor one state over the other, such that $\langle \tau_{\rm P,AP} \rangle = \tau_{0} \exp\left[ \Delta E \left(1 \pm I/I_{C}\right) / (k_{B}T) \right]$, where $\Delta E$ is the energy barrier between P and AP states at zero bias, and $I_{C}$ is the critical switching current at zero temperature~\cite{PRB_69_Li_2004,PRB_84_Rippard_2011}. In our convention, a positive current tends to stabilize the (AP) magnetization state. 

We studied the response of the SP-MTJ to a weak oscillating excitation current. Under zero bias field~\footnote{For the sake of clarity,  \emph{zero bias field} refers to the situation where the external field exactly compensates the dipolar influence of the SAF on the free layer, so that both P and AP states show equal stabilities at zero current}, we injected a square wave periodic current of different amplitudes, $I_{\rm ac}=250, 200, 150$ and 100 $\mu$A, and frequencies between $5$ Hz and $2$ kHz, and we monitored the voltage across the junction with an oscilloscope. For each ac-current amplitude $I_{\rm ac}$ and frequency $F_{\rm ac}$ considered, 10 traces of 2 seconds in duration were recorded in order to obtain sufficient statistics on the response of the stochastic device. By analysing the voltage data, we constructed the time evolution of the free layer magnetization and examined how it was influenced by the different parameters of the input ac current.

\begin{figure}
	\centering
		\includegraphics[width=\linewidth]{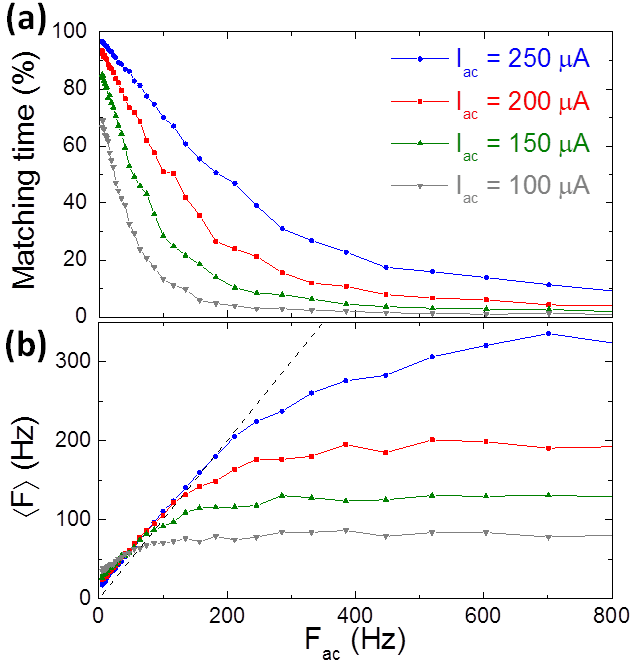}
		\caption{(Color online). MTJ response to different current amplitudes $I_{\rm ac}=250, 200, 150$ and $100\mu$A. (a) Matching time between the resistance response and the excitation signal. (b) Average oscillation frequency of the resistance response $\langle F \rangle$ versus excitation frequency $F_{\rm ac}$. The dashed line corresponds to a match between response and excitation frequencies.}
	\label{fig:FresVsFcurr}
\end{figure}

\begin{figure*}[t]
	\centering
	\includegraphics[width=\linewidth]{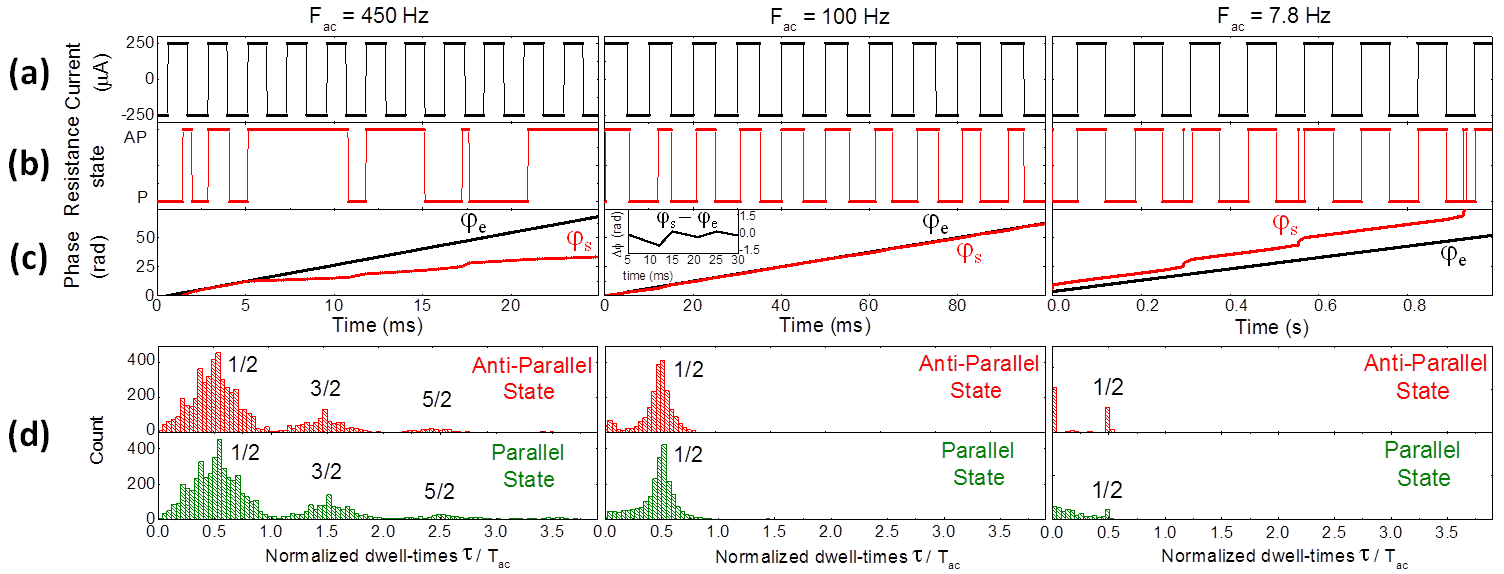}
	\caption{(Color online) SP-MTJ response to square wave current excitations with $250$ $\mu$A amplitude and frequencies $F_{\rm ac}=450$ Hz (left), $F_{\rm ac}=100$ Hz (center), $F_{\rm ac}=7.8$ Hz (right). Sample of the temporal evolution of (a) the current through the MTJ, (b) its resistance response, and (c) the piecewise linear reconstructed phase for the current ($\varphi_{e}$, black) and for the resistance/magnetization ($\varphi_{s}$, red). (d) Associated dwell-time distributions for both high (AP) and low (P) resistance states.}
	\label{fig:TracesAndDistributions}
\end{figure*}

A first method to quantify the correlation between the stochastic oscillator and the excitation source involves the percentage of time during which the two signals are in phase, which we denote as the ``matching time''. Fig.~\ref{fig:FresVsFcurr}a shows that this matching time increases monotonously as the excitation frequency $F_{\rm ac}$ decreases. This feature is characteristic of stochastic resonance~\cite{RMP_70_Gammaitoni_1998} and has already been observed in other studies~\cite{PRL_105_Cheng_2010, APL_103_Cheng_2013, PRB_83_Finocchio_2011}. When the driving frequency is large compared to the natural oscillation rate of the stochastic oscillator, the latter does not respond to the input signal, and the matching time is small. By lowering the input frequency, the ability of the oscillator to adjust to the external forcing increases, and the matching time also increases accordingly. As Fig.~\ref{fig:FresVsFcurr}a shows, the matching time also increases with increasing amplitude of the ac current, reaching values of 96.8 $\%$ at 5 Hz and $I_{\rm ac}=250$ $\mu$A.

However, the matching time does not provide any information regarding the influence that the excitation current has on the frequency of the stochastic oscillator. To investigate this point, we also examined the evolution of the mean oscillation frequency $\langle F \rangle$ as a function of the frequency of the excitation current, $F_{\rm ac}$ (Fig.~\ref{fig:FresVsFcurr}b). When the frequency of the input signal is too large for the magnetization to follow, the mean frequency of the stochastic oscillator plateaus to a constant value that is determined by the amplitude of the ac current. When the excitation frequency is reduced, on the other hand, we observe a clear pulling of the mean frequency of the stochastic oscillator toward the excitation frequency. This pulling effect becomes more efficient as the amplitude of the excitation increases. However, as Fig.~\ref{fig:FresVsFcurr}b shows, the mean frequency of the stochastic oscillator deviates from the frequency of the input signal (dashed line) at low frequencies at which the matching time percentage increases.

In order to get a better insight into this effect, we examined the time traces of the SP-MTJ response for different frequencies of the excitation current. The square wave current of amplitude $I_{\rm ac}=250$ $\mu$A and the temporal resistance evolution of the MTJ are shown  in Fig.~\ref{fig:TracesAndDistributions}a and \ref{fig:TracesAndDistributions}b, respectively. We have also reconstructed the piecewise linear phases~\cite{CWN_13_Freund_2003} associated with both the current ($\varphi_{e}$) and magnetization ($\varphi_{s}$) oscillations (Fig.~\ref{fig:TracesAndDistributions}c) and extracted the dwell-time distributions for both P and AP states (Fig.~\ref{fig:TracesAndDistributions}d).

Let us first consider the case of a high excitation frequency $F_{\rm ac}=450$ Hz compared to the natural frequency $\langle F \rangle$ of the oscillator (Fig.~\ref{fig:TracesAndDistributions}, left column). As the resistance time trace shows, the magnetization switching is largely correlated with the reversals of the current polarity  (Fig.~\ref{fig:TracesAndDistributions}a, b). A direct consequence is that the dwell-times no longer follow a Poisson distribution, as observed when a constant current is applied, but rather exhibit peaks around $(n + 1/2) T_{\rm ac}$ where $T_{\rm ac}$ is the period of the excitation (Fig.~\ref{fig:TracesAndDistributions}c). Such correlations can be explained as follows. A dwell-time close to $T_{\rm ac}/2$ means that two consecutive ac-current polarity reversals both induce a magnetization reversal. If the system does not follow a polarity reversal, it has to wait for one more period for the next reversal ($\tau \approx 3T_{\rm ac}/2$), or two more periods ($\tau \approx 5T_{\rm ac}/2$), and so on. Nevertheless, the stochastic nature of the switching still manifests itself in the Poisson-like decay in the amplitude of the peaks in the dwell time distribution as $\tau$ increases. While the MTJ responds resonantly under the influence of the excitation, due to the phenomenon of stochastic resonance, synchronization to the external signal is not achieved. The mean oscillation frequency of the stochastic oscillator $\langle F \rangle=282$ Hz (Fig.~\ref{fig:FresVsFcurr}b) remains significantly lower than the excitation frequency $F_{\rm ac}$ and the matching time between the two signals is low at $17.4\%$ (Fig.~\ref{fig:FresVsFcurr}a). This can also be seen in the piecewise linear phase evolution, where the phase of the stochastic oscillator increases more slowly than the phase of the input square wave, with no visible correlation between the two (Fig.~\ref{fig:TracesAndDistributions}d).

\begin{figure}
	\centering
		\includegraphics[width=\linewidth]{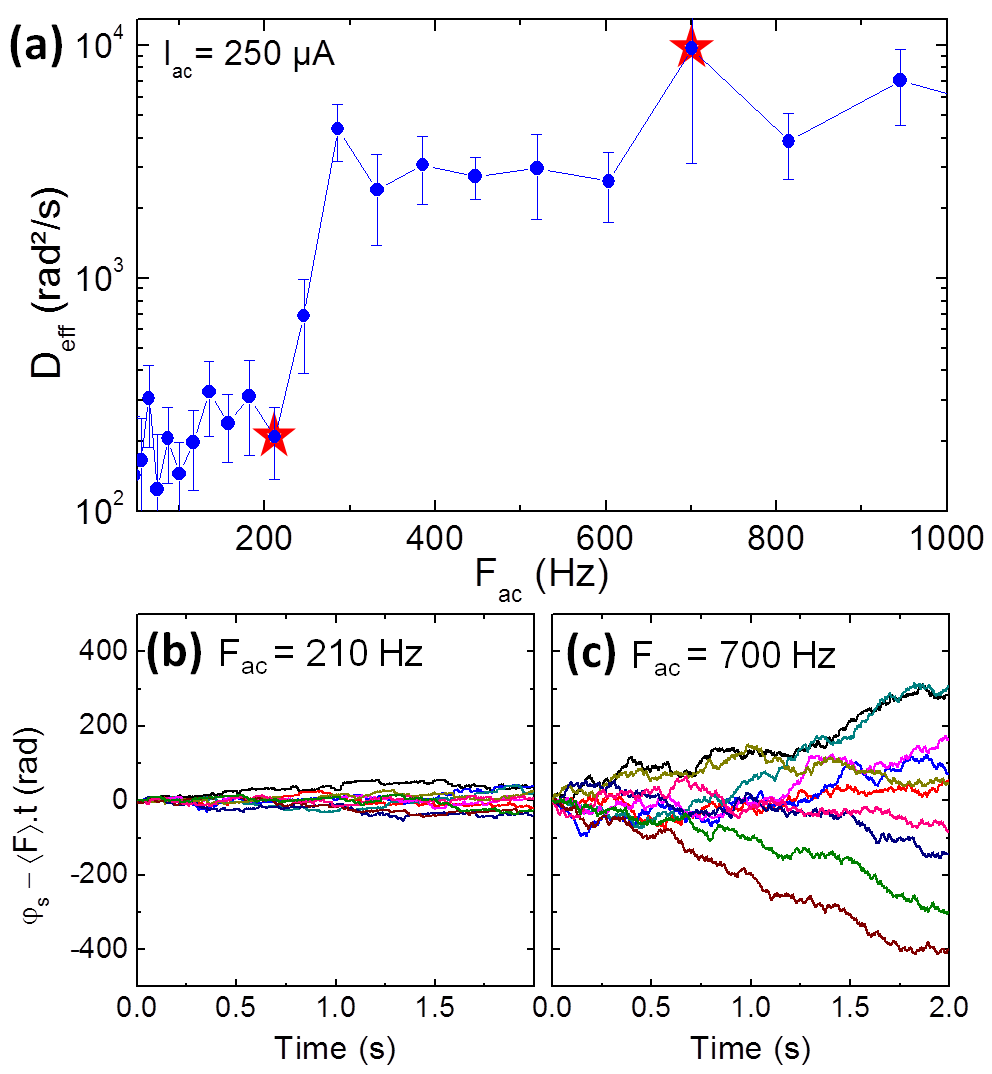}
		\caption{(Color online) (a) Effective diffusion constant $D_{\rm eff}$ as a function of the input excitation frequency $F_{\rm ac}$ for a current amplitude $I_{\rm ac}=250$ $\mu$A. Ten time traces of the stochastic oscillator phase for input frequencies of (b) $210$ Hz and (c) $700$ Hz, indicated by the red stars in (a). The linear variation corresponding to the average frequency was subtracted for clarity.}
	\label{fig:DeffVsFac}
\end{figure}

As the excitation frequency is reduced to $F_{\rm ac}=100$ Hz, the response of the system changes and the stochastic nature of the oscillator becomes less apparent (Fig.~\ref{fig:TracesAndDistributions}, center column). As the resistance time traces show, the magnetization switches with each reversal of the current polarity and the dwell-time distributions exhibit only one peak around $T_{\rm ac}/2$. In addition, we note almost no occurrence during the 20 second measurement of a dwell-time superior to $T_{\rm ac}$, which would indicate one missed oscillation or phase slip. Furthermore, the reconstructed phase of the stochastic oscillator exhibits a quasi-linear variation and follows closely the phase of the input signal. As such, the stochastic behavior is considerably reduced and is only observed as small, bounded fluctuations of the piecewise linear phase-shift $\Delta\phi$ around zero (see inset of Fig.~\ref{fig:TracesAndDistributions}c, center). A near perfect phase synchronization is therefore achieved during this time interval for low current densities, since 250 $\mu$A in our junctions corresponds to a  density below $3 \times 10^6$ A/cm$^2$. The mean frequency of the stochastic oscillator is measured to be $\langle F \rangle=111$ Hz (Fig.~\ref{fig:FresVsFcurr}b) and the matching percentage has increased to $70.0\%$ (Fig.~\ref{fig:FresVsFcurr}a). 

As the excitation frequency is further reduced to $F_{\rm ac}=7.8$ Hz, we observe the appearance of a large number of ``glitches'' in the magnetization switching (Fig.~\ref{fig:TracesAndDistributions}, right column). Since the current remains in each polarity for a long time compared to the natural dwell-times, the probability of thermally-driven back-and-forth switching of the magnetization increases, which gives rise to the observed glitches. These events occur over a short time scale compared to the period of the input square wave current and appear as localized $2\pi$ jumps in the phase of the stochastic oscillator. Outside of these glitches, the magnetization is seen to switch with each reversal of the current polarity, which results in a quasi-linear variation of the stochastic oscillator phase. While the matching percentage reaches a very high value of $95.9\%$, the presence of the glitches lead to an increase in the mean frequency of the stochastic oscillator oscillation, $\langle F \rangle=20.1$ Hz, which is significantly larger than the excitation frequency.

The broad features of the variation of the dwell-time distribution with the frequency of the input forcing signal described above are consistent with previous theoretical and experimental work on periodically-forced bistable systems. While phase locking of stochastic oscillators has been studied previously~\cite{PRE_58_Neiman_1998,CWN_13_Freund_2003}, the existence of glitches in the synchronized state and their role in frequency mismatches have not been considered to the best of our knowledge. In fact, these glitches and the associated phase jumps are very similar in nature to the thermally induced phase slips that can be observed when noisy oscillators are phase locked. To quantify such phenomena, we analyze the phase evolution of the stochastic magnetic oscillator as a one dimensional random walk. Using the measured time traces, we can derive the diffusion constant for the phase~\cite{PRE_58_Neiman_1998,CWN_13_Freund_2003},
\begin{equation}
	D_{\rm eff} = \frac{1}{2} \frac{d}{dt} \left( \langle {\varphi_{s,i}}^{2}(t)\rangle_{i} - \langle \varphi_{s,i}(t) \rangle_{i}^{2} \right),
\end{equation}
which quantifies the randomness of the evolution of magnetization oscillations, and $D_{\rm eff}=0$ would correspond to a deterministic (e.g. perfectly harmonic) behaviour of the oscillator. Here, the index $i\in[1:10]$ corresponds to the measurement number. The evolution of the diffusion coefficient versus excitation frequency at constant current amplitude is shown in Fig.~\ref{fig:DeffVsFac}a for $I_{\rm ac}=250$ $\mu$A. We note a drastic transition from the random to the synchronized regime starting at a critical frequency $F_{C}\approx300$ Hz, which is very close to the natural frequency of the stochastic oscillator for this input current amplitude (320 Hz, as shown in Fig.~\ref{fig:FresVsFcurr}b). The system indeed switches from a high excitation frequency regime where the diffusion constant is large, corresponding to the free running stochastic oscillator, to a regime at low excitation frequencies where the oscillator is entrained by the ac-current input signal. This result shows that despite the glitches and the difference between the average oscillation rate and the excitation frequency, the stochastic oscillator falls into a synchronized regime in which it is phase-locked to the excitation. To illustrate the difference in the dispersion below and above the critical frequency, we present in Fig.~\ref{fig:DeffVsFac}b and \ref{fig:DeffVsFac}c the 10 $\varphi_{s}(t)$ measurements for $F_{\rm ac}=210$ Hz and $F_{\rm ac}=700$ Hz, respectively, on the same scale. We can clearly see that for $F_{\rm ac}>F_{C}$ the dispersion between oscillators is very high, while the dispersion is strongly suppressed by external excitation for $F_{\rm ac}<F_{C}$. The non-vanishing $D_{\rm eff}$ in the synchronized regime represents the remaining stochastic behavior associated with the random occurrence of glitches and thermal noise.

In conclusion, we have demonstrated that a stochastic magnetic oscillator can phase lock to an input signal with a frequency lower than the natural frequency of the oscillator. While the synchronization is not perfect due to unavoidable phase slips, it occurs for values of the forcing current that are below the critical values required for deterministic switching at zero temperature. This system is therefore extremely promising for applications where low energy is crucial and thermal noise has to be leveraged, such as bio-inspired associative-memories based on spin torque nano-oscillator networks~\cite{TIWCN__Csaba_2012}. 

\subsection*{Acknowledgements}
The authors acknowledge financial support from the FET-OPEN project Bambi. A. Accioly acknowledges financial support from the Brazilian agency National Council for Scientic and Technological Development (CNPq, Project No. 245555/2012-9), and A. Mizrahi from the Ile-de-France regional government through the DIM nano-K program. 

\bibliography{StochasticSync_v2}

\end{document}